# A semirelativistic treatment of spinless particles subject to the Yukawa potential with arbitrary angular momenta

Majid Hamzavi[1*], Sameer M. Ikhdair[2**], M. Solaimani[1]

[1]*Department of Basic Sciences, Shahrood Branch, Islamic Azad University, Shahrood, Iran*

[2]*Physics Department, Near East University, Nicosia, North Cyprus, Mersin 10, Turkey*

[*]Corresponding author: *majid.hamzavi@gmail.com*

[**] *sikhdair@neu.edu.tr*

**Abstract**

We obtain analytical solutions of the two-body spinless Salpeter (SS) equation with the Yukawa potential within the conventional approximation scheme to the centrifugal term for any $l$-state. The semi-relativistic bound state energy spectra and the corresponding normalized wave functions are calculated by means of the Nikiforov-Uvarov (NU) method. We also obtain the numerical energy spectrum of the SS equation without any approximation to centrifugal term for the same potential and compare them with the approximated numerical ones obtained from the analytical expressions. It is found that the exact numerical results are in good agreement with the approximated ones for the lower energy states. Special cases are treated like the nonrelativistic limit and the solution for the Coulomb problem.

## 1. Introduction

The Bethe–Salpeter equation [1], named after Hans Bethe and Edwin Salpeter, describes the bound states of a two-body (particles) quantum field theoretical system in a relativistic covariant formalism. The equation was actually first published in 1950 at the end of a paper by Yoichiro Nambu but without derivation [2].

Due to its generality and its application in many branches of theoretical physics, the Bethe–Salpeter equation appears in many different forms [3-10]. Different aspects of this equation have been elegantly studied mainly by Lucha *et al* in [11-24] and some other authors where they have worked on many interesting approaches to deal with its nonlocal Hamiltonian [25-32].

The Yukawa potential or static screening Coulomb potential (SSCP) is given by

$$V(r) = -V_0 \frac{e^{-ar}}{r}, \tag{1}$$

where $V_0 = \alpha Z$, $\alpha = 1/137.037$ is the fine-structure constant and $Z$ is the atomic number and $a$ is the screening parameter. This potential is often used to compute bound-state normalizations and energy levels of neutral atoms [33-36] which have been studied over the past years. It is known that SSCP yields reasonable results only for the innermost states when $Z$ is large. However, for the outermost and middle atomic states, it gives rather poor results. The bound-state energies of the SSCP with $Z = 1$ have been studied in the light of the shifted large-$N$ expansion method [37]. Ikhdair and Sever investigated energy levels of neutral atoms by applying an alternative perturbative scheme in solving the Schrödinger equation for the Yukawa potential model with a modified screening parameter [36]. They studied bound states of the Hellmann potential, which represents the superposition of the attractive Coulomb potential $-a/r$ and the Yukawa potential $b\exp(-\delta r)/r$ of arbitrary strength $b$ and the screening parameter $\delta$ [38]. They also considered the bound states of the exponential cosine-screened Coulomb potential [39] and a more general exponential screened Coulomb (MGESC) potential [40].

Recently, Karakoc and Boztosun applied the asymptotic iteration method to solve the radial Schrödinger equation for the Yukawa type potentials [41]. Further, Gönül *et al* proposed a new approximation scheme to obtain analytic expressions for the bound state energies and wave functions of Yukawa like potentials [42].

Very recently, Ikhdair has approximately solved the relativistic Dirac equation with the Yukawa potential for any spin-orbit quantum number $\kappa$ in the presence of spin and pseudospin symmetry [43]. In addition, Liverts *et al* used the quasi-linearization method (QLM) for solving the Schrödinger equation with Yukawa potential [44].

The aim of the present work is to study the Yukawa potential within the framework of a semi-relativistic SS equation. First, we approximately calculate the bound-state energy eigenvalues and their corresponding normalized wave functions with arbitrary $l$-states for the short-range Yukawa potential by using a parametric generalization of the NU method [45]. This shortcut is a powerful tool in solving second-order differential equation and has proved its effectiveness in solving various potential models over the past few years. Second, we are going to compare our approximated numerical energy eigenvalues with the exact numerical ones calculated by solving the same quantum system without making approximation to the centrifugal term.

The structure of the paper is as follows. In Section 2, we review the SS equation and apply it to the Yukawa potential interaction to obtain the semirelativistic SS bound-state energy spectrum and their corresponding normalized wave functions for two-interacting particles. In section 2, we reduce our solution into the nonrelativistic limit and obtain the Schrödinger bound-states of the energy spectrum and normalized wave functions. Further, we obtain the non-relativistic solution of the Coulomb field. Finally, in Section 3 we give our final comments and conclusion.

## 2. Spinless Salpeter equation and application to Yukawa potential

The spinless Salpeter (SS) equation for a two-body system under a spherically symmetric potential in the center-of-mass system has the form [25,26,30, 31]

$$\left[\sum_{i=1,2}\sqrt{-\Delta + m_i^2} + V(r) - M\right]\chi(\vec{r}) = 0, \qquad \Delta = \nabla^2, \tag{2}$$

where the kinetic energy terms involving the operation $\sqrt{-\Delta + m_i^2}$ are nonlocal operators and $\chi(\vec{r}) = R_{nl}(r)Y_{lm}(\theta,\phi)$, where $R_{nl}(r) = r^{-1}\psi_{nl}(r)$, stands for the semi-relativistic wave function. For heavy interacting particles, the kinetic energy operators in Eq. (2) can be approximated as [2-32]

$$\sqrt{-\Delta + m_i^2} = m_1 + m_2 - \frac{\Delta}{2\mu} - \frac{\Delta^2}{8\eta^3} - ..., \tag{3}$$

where $\mu = m_1 m_2/m_1 + m_2$ stands for the reduced mass and $\eta = \mu\left(m_1 m_2/(m_1 m_2 - 3\mu^2)\right)^{1/3}$ is an introduced useful mass parameter [25,26]. The above Hamiltonian containing relativistic corrections up to order $(v^2/c^2)$ and is called a generalized Breit-Fermi Hamiltonian [11-15]. Using an appropriate transformation and following the procedures explained in Ref. [30] (see Eqs. (13)-(18)), one can then recast the semirelativistic SS equation as

$$\left[-\frac{\hbar^2}{2\mu}\frac{d^2}{dr^2} + \frac{l(l+1)\hbar^2}{2\mu r^2} + W_{nl}(r) - \frac{W_{nl}^2(r)}{2\tilde{m}}\right]\psi_{nl}(r) = 0, \tag{4}$$

where

$$W_{nl}(r) = V(r) - E_{nl}, \text{ and } \tilde{m} = \frac{\eta^3}{\mu^2} = \frac{m_1 m_2 \mu}{m_1 m_2 - 3\mu^2}. \tag{5}$$

Now, we intend to solve the above semi-relativistic SS equation with the Yukawa potential interaction (1). Thus, the insertion of Eq. (1) into Eq. (5) allows one to obtain

$$\left[\frac{d^2}{dr^2} - \frac{l(l+1)}{r^2} + \frac{2\mu}{\hbar^2}\left(V_0 \frac{e^{-ar}}{r} + E_{nl} + \frac{V_0^2}{2\tilde{m}}\frac{e^{-2ar}}{r^2} + \frac{E_{nl}V_0}{\tilde{m}}\frac{e^{-ar}}{r} + \frac{E_{nl}^2}{2\tilde{m}}\right)\right]\psi_{nl}(r) = 0. \tag{6}$$

Since Eq. (6) does not admit exact analytical solution due to the presence of the strong singular centrifugal term $r^{-2}$, we resort to use a proper approximation to deal with this term. So, we employt he conventional approximation scheme introduced by Greene and Aldrich [46]:

$$\frac{1}{r^2} \approx 4a^2 \frac{e^{-2ar}}{(1-e^{-2ar})^2}, \tag{7a}$$

$$\frac{1}{r} \approx 2a \frac{e^{-ar}}{(1-e^{-2ar})}, \tag{7b}$$

which is valid only for a short-range potential, i.e., $ar \ll 1$ and used only to calculate the lowest energy states as mentioned in [47]. Therefore, to see the accuracy of our approximation, we plot the Yukawa potential (1) and its approximation [43]

$$V(r) = -2aV_0 \frac{e^{-2ar}}{\left(1-e^{-2ar}\right)}, \tag{8}$$

with parameters $V_0 = 1.0$ and $a = 0.01 fm^{-1}$, as shown in Figure 1. Thus, the approximate analytical solution of the SS equation with the Yukawa potential can be obtained by inserting Eq. (7) into Eq. (6) as

$$\left[\frac{d^2}{dr^2} - 4a^2 \frac{l(l+1)e^{-2ar}}{(1-e^{-2ar})^2}\right.$$
$$\left. + \frac{2\mu}{\hbar^2}\left(2aV_0 \frac{e^{-2ar}}{(1-e^{-2ar})} + E_{nl} + \frac{2a^2V_0^2}{\tilde{m}}\frac{e^{-4ar}}{(1-e^{-2ar})^2} + \frac{2aE_{nl}V_0}{\tilde{m}}\frac{e^{-2ar}}{(1-e^{-2ar})} + \frac{E_{nl}^2}{2\tilde{m}}\right)\right]\psi_{nl}(r) = 0. \tag{9}$$

Now, we can recast the above equation into a simpler form which is amendable to the NU [45] solution by making the change of variable $s = e^{-2ar}$ to obtain

$$\left[\frac{d^2}{ds^2} + \frac{1-s}{s(1-s)}\frac{d}{ds} + \frac{1}{s^2(1-s)^2}(-As^2 + Bs - C)\right]\psi_{nl}(s) = 0, \tag{10}$$

where

$$A = \frac{\mu}{\hbar^2}\left(V_0 - \frac{E_{nl}}{2a}\right)\left[\frac{1}{a} - \frac{1}{\tilde{m}}\left(V_0 - \frac{E_{nl}}{2a}\right)\right], \quad (11a)$$

$$B = -l(l+1) + \frac{\mu}{\hbar^2 a}\left[\left(V_0 - \frac{E_{nl}}{a}\right) + \frac{E_{nl}}{\tilde{m}}\left(V_0 - \frac{E_{nl}}{2a}\right)\right], \quad (11b)$$

$$C = -\frac{\mu}{\hbar^2}\frac{E_{nl}}{2a^2}\left(1 + \frac{E_{nl}}{2\tilde{m}}\right). \quad (11c)$$

In Appendix A, we present our shortcut to the NU method used to obtain the analytical solution of Eq. (10). Hence, comparing equation (10) with (A1), we find the parametric coefficients:

$$c_1 = 1, \qquad c_2 = 1, \qquad c_3 = 1. \quad (12)$$

In addition, the relation (A5) determines the rest of the coefficients as

$$c_4 = 0, \qquad c_5 = -\frac{1}{2},$$

$$c_6 = \frac{1}{4} + A, \qquad c_7 = -B,$$

$$c_8 = C, \qquad c_9 = \frac{1}{4} + A - B + C = -\frac{\mu V_0^2}{\hbar^2 \tilde{m}} + \left(l + \frac{1}{2}\right)^2,$$

$$c_{10} = 2\sqrt{C}, \qquad c_{11} = \sqrt{1 + 4(A - B + C)},$$

$$c_{12} = \sqrt{C}, \qquad c_{13} = \frac{1}{2}\left[1 + \sqrt{1 + 4(A - B + C)}\right]. \quad (13)$$

Therefore, from relation (A10), we can obtain the binding energy equation as

$$\sqrt{\frac{\mu}{\hbar^2}\left[\frac{1}{a}\left(V_0 - \frac{E_{nl}}{2a}\right) - \frac{1}{\tilde{m}}\left(\frac{E_{nl}}{2a} - V_0\right)^2\right]} - \sqrt{-\frac{\mu}{\hbar^2}\frac{E_{nl}}{2a^2}\left(V_0 + \frac{E_{nl}}{2\tilde{m}}\right)} = n + \nu + \frac{1}{2}, \quad (14a)$$

$$\nu = \sqrt{\left(l + \frac{1}{2}\right)^2 - \frac{\mu}{\hbar^2}\frac{V_0^2}{\tilde{m}}}, \quad (14b)$$

where $\nu$ is a new quantum number and the binding energy solution is chosen to be negative (i.e., $E_{nl} < 0$) and small satisfying $E_{nl} \ll \tilde{m}$ [25,26]. Equation (14a) appears as a complicated transcendental energy equation acquires two solutions. However, we select the negative binding energy as a physical solution to the transcendental equation.

To show the accuracy of this approximation scheme, we have calculated the exact numerical energy eigenvalues of Eq. (6) without making approximation to the centrifugal term by using the finite difference method [48-50] with arbitrary radial and orbital quantum numbers $n$ and $l$ for different values of the screening parameter $a$. The results are given in Table 1 and compared with those obtained by using Eq. (14). In the low screening region, for various $(n,l)$ states when the screening parameter $a = 0.001\, fm^{-1}$, we have found the percentage error $\left|\frac{E(approx) - E(num)}{E(num)}\right|\%$ for a few lowest states: $0.0454\,\%$; $3.403\,\%$ and $0.866\,\%$; $6.478\,\%$, $0.936\,\%$ and $1.292\,\%$ for $(1,0)$; $(2,0)$ and $(2,1)$; $(3,0)$, $(3,1)$ and $(3,2)$, respectively. However, the percentage error increases with the increasing of the screening parameter, $a = 0.01\, fm^{-1}$, it becomes $0.269\,\%$; $3.598\,\%$ and $1.238\,\%$; $7.206\,\%$, $0.596\,\%$ and $0.150\,\%$. We take the following set of physical parameter values, $m_1 = 5\, fm^{-1}$, $m_2 = 5\, fm^{-1}$, $V_0 = 1$ and $a = (0.01,\ 0.005,\ 0.001)\, fm^{-1}$. It is found that the numerical energy states obtained by means of Eq. (6) are in good agreement with the ones obtained by Eq. (14) for low screening regime, i.e., $a \ll 1$ since the approximation works well only for the lowest states [47].

Now, we seek to find the corresponding wave functions, referring to Eq. (13) and relations (A11) and (A12) of Appendix A, we find the functions

$$\rho(s) = s^{2\varepsilon_{nl}}(1-s)^{2\nu}, \tag{15a}$$

$$\phi(s) = s^{\varepsilon_{nl}}(1-s)^{\frac{1}{2}+\nu}, \tag{15b}$$

where

$$\varepsilon_{nl} = \sqrt{-\frac{\mu}{2\hbar^2}\frac{E_{nl}}{a^2}\left(1+\frac{E_{nl}}{2\tilde{m}}\right)}. \tag{16}$$

Hence, relation (A13) gives

$$y_n(s) = P_n^{(2\varepsilon_{nl},2\nu)}(1-2s). \tag{17}$$

By using $R_{nl}(s) = \phi(s)y_n(s)$, we get the radial SS wave functions for the Yukawa potential from relation (A14) as

$$\psi_{nl}(s) = A_{nl} s^{\varepsilon_{nl}}(1-s)^{\frac{1}{2}+\nu} P_n^{(2\varepsilon_{nl},2\nu)}(1-2s) \tag{18}$$

and upon substituting $s = e^{-2ar}$, we obtain

$$\psi_{nl}(r) = A_{nl}\left(e^{-2ar}\right)^{\varepsilon_{nl}}(1-e^{-2ar})^{\frac{1}{2}+\nu}P_n^{(2\varepsilon_{nl},2\nu)}(1-2e^{-2ar}), \qquad (19)$$

where the normalization constant $A_{nl}$ is calculated in details in Appendix B as

$$A_{nl} = \sqrt{\frac{2a\varepsilon_{nl}n!(2n+2\nu+2\varepsilon_{nl}+1)\Gamma(n+2\nu+2\varepsilon_{nl}+1)}{(n+\nu+1/2)\Gamma(n+2\nu+1)\Gamma(n+2\varepsilon_{nl}+1)}}. \qquad (20)$$

The new quantum number $\nu$ and the energy parameter $\varepsilon_{nl}$ appearing in Eq. (19) are defined in Eqs. (14b) and (16), respectively. To show the accuracy of our approximation, we plot in Figure 2 the approximated and exact numerical wave functions of the SS equation with the Yukawa potential for $n=1$ and $l=0$ state.

## 3. A Few Special Cases

First, we consider the nonrelativistic limiting case when $W_{nl}(r) \ll \tilde{m}$, then $W_{nl}^2(r) \approx 0$, from which Eq. (4) can be recast in the form of Schrödinger equation having the energy formula and wave functions obtained from Eqs. (14) and (19) as

$$E_{nl} = -a\left[\frac{\hbar^2 a}{2\mu}(n+l+1)^2 + \frac{\mu}{2\hbar^2 a}\frac{V_0^2}{(n+l+1)^2} - 2V_0\right], \qquad (21)$$

and

$$\psi_{nl}(r) = C_{nl}\left(e^{-2ar}\right)^{\lambda_{nl}}(1-e^{-2ar})^{l+1}P_n^{(2\lambda_{nl},2l+1)}(1-2e^{-2ar}), \quad \lambda_{nl} = \frac{1}{a}\sqrt{-\frac{\mu E_{nl}}{2\hbar^2}}, \qquad (22)$$

respectively. The Schrödinger normalization constant $C_{nl}$ can be easily found from Appendix B as

$$C_{nl} = \sqrt{\frac{4a\lambda_{nl}n!(n+l+\lambda_{nl}+1)\Gamma(n+2l+2\lambda_{nl}+2)}{(n+l+1)\Gamma(n+2l+2)\Gamma(n+2\lambda_{nl}+1)}}. \qquad (23)$$

In Table 2, we calculate some approximation and numerical energy eigenvalues of the Schrödinger equation with the Yukawa potential for arbitrary values of radial and orbital $n$ and $l$ quantum numbers, respectively. In our numerical calculations, we used the parameter values $m_1 = 5\,fm^{-1}$, $m_2 = 5\,fm^{-1}$, $V_0 = 1$ and $a = (0.01, 0.005, 0.001)\,fm^{-1}$.

Second, upon inserting $a=0$ in Eq. (21), the bound-state energy formula of two particles interacting via the Coulomb potential within the Schrödinger equation is

$$E_{nl} = -\frac{\mu}{2\hbar^2} \frac{V_0^2}{(n+l+1)^2}. \tag{24}$$

## 4. Final remarks and conclusion

We have presented approximate analytical bound state solutions of the two body SS equation with the Yukawa potential for arbitrary $l$-states by considering an approximation to the centrifugal term. The semirelativistic bound-state energy eigenvalues and their corresponding normalized wave functions are obtained by the parametric generalization of the NU method. It is found that wave function solution can be expressed in terms of the Jacobi polynomials. The results obtained from exact numerical solution of Eq. (6) are in good agreement with the approximate solution of the quantum system in Eq. (14) for low screening region (short-range potential).

The aim of solving the Yukawa potential returns to the following reasons: First, in the low screening region where the screening parameter $a$ is small (i.e., $a \ll 1$), the potential reduces to the Killingbeck potential [51-53], i.e., $V(r) = ar^2 + br - c/r$, where $a$, $b$ and $c$ are potential constants that can be obtained after making expansion to the Yukawa potential. Second, it can also be reduced into the Cornell potential [54,55], i.e., $V(r) = br - c/r$. These two potentials are usually used in the study of mesons and baryons. Third, when the screening parameter approaches to zero, the Yukawa potential turns to become the Coulomb potential.

## 4. Acknowledgements

S. M. Ikhdair acknowledges the partial support provided by the Scientific and Technological Research Council of Turkey (TÜBİTAK). The authors also thank the kind referee for the helpful comments and suggestions which have improved the manuscript greatly.

**Appendix A: Parametric Generalization of the NU method**

The NU method is used to solve second order differential equations with an appropriate coordinate transformation $s = s(r)$ [45]

$$\psi_n''(s) + \frac{\tilde{\tau}(s)}{\sigma(s)}\psi_n'(s) + \frac{\tilde{\sigma}(s)}{\sigma^2(s)}\psi_n(s) = 0, \qquad (A1)$$

where $\sigma(s)$ and $\tilde{\sigma}(s)$ are polynomials, at most of second degree, and $\tilde{\tau}(s)$ is a first-degree polynomial. To make the application of the NU method simpler and direct without need to check the validity of solution. We present a shortcut for the method. So, at first we write the general form of the Schrödinger-like equation (A1) in a more general form applicable to any potential as follows [56-58]

$$\psi_n''(s) + \left(\frac{c_1 - c_2 s}{s(1 - c_3 s)}\right)\psi_n'(s) + \left(\frac{-As^2 + Bs - C}{s^2(1 - c_3 s)^2}\right)\psi_n(s) = 0, \qquad (A2)$$

satisfying the wave functions

$$\psi_n(s) = \phi(s) y_n(s). \qquad (A3)$$

Comparing (B2) with its counterpart (B1), we obtain the following identifications:

$$\tilde{\tau}(s) = c_1 - c_2 s, \quad \sigma(s) = s(1 - c_3 s), \quad \tilde{\sigma}(s) = -As^2 + Bs - C. \qquad (A4)$$

Following the NU method [45], we obtain the followings [56-58],

(i) the relevant constant:

$$c_4 = \frac{1}{2}(1 - c_1), \qquad c_5 = \frac{1}{2}(c_2 - 2c_3),$$

$$c_6 = c_5^2 + A, \qquad c_7 = 2c_4 c_5 - B,$$

$$c_8 = c_4^2 + C, \qquad c_9 = c_3(c_7 + c_3 c_8) + c_6,$$

$$c_{10} = c_1 + 2c_4 + 2\sqrt{c_8} - 1 > -1, \qquad c_{11} = 1 - c_1 - 2c_4 + \frac{2}{c_3}\sqrt{c_9} > -1, \; c_3 \neq 0,$$

$$c_{12} = c_4 + \sqrt{c_8} > 0, \qquad c_{13} = -c_4 + \frac{1}{c_3}(\sqrt{c_9} - c_5) > 0, \; c_3 \neq 0. \qquad (A5)$$

(ii) the essential polynomial functions:

$$\pi(s) = c_4 + c_5 s - \left[ \left( \sqrt{c_9} + c_3 \sqrt{c_8} \right) s - \sqrt{c_8} \right], \tag{A6}$$

$$k = -(c_7 + 2c_3 c_8) - 2\sqrt{c_8 c_9}, \tag{A7}$$

$$\tau(s) = c_1 + 2c_4 - (c_2 - 2c_5)s - 2\left[ \left( \sqrt{c_9} + c_3 \sqrt{c_8} \right) s - \sqrt{c_8} \right], \tag{A8}$$

$$\tau'(s) = -2c_3 - 2\left( \sqrt{c_9} + c_3 \sqrt{c_8} \right) < 0. \tag{A9}$$

(iii) The energy equation:

$$c_2 n - (2n+1)c_5 + (2n+1)\left( \sqrt{c_9} + c_3 \sqrt{c_8} \right) + n(n-1)c_3 + c_7 + 2c_3 c_8 + 2\sqrt{c_8 c_9} = 0. \tag{A10}$$

(iv) The wave functions

$$\rho(s) = s^{c_{10}} (1 - c_3 s)^{c_{11}}, \tag{A11}$$

$$\phi(s) = s^{c_{12}} (1 - c_3 s)^{c_{13}}, \quad c_{12} > 0, \quad c_{13} > 0, \tag{A12}$$

$$y_n(s) = P_n^{(c_{10}, c_{11})}(1 - 2c_3 s), \quad c_{10} > -1, \quad c_{11} > -1, \tag{A13}$$

$$\psi_{n\kappa}(s) = N_{n\kappa} s^{c_{12}} (1 - c_3 s)^{c_{13}} P_n^{(c_{10}, c_{11})}(1 - 2c_3 s). \tag{A14}$$

where $P_n^{(\mu,\nu)}(x)$, $\mu > -1$, $\nu > -1$, and $x \in [-1,1]$ are Jacobi polynomials with

$$P_n^{(\alpha,\beta)}(1 - 2s) = \frac{(\alpha+1)_n}{n!} {}_2F_1(-n, 1+\alpha+\beta+n; \alpha+1; s), \tag{A15}$$

and $N_{n\kappa}$ is a normalization constant. Also, the above wave functions can be expressed in terms of the hypergeometric function as

$$\psi_{n\kappa}(s) = N_{n\kappa} s^{c_{12}} (1 - c_3 s)^{c_{13}} {}_2F_1(-n, 1+c_{10}+c_{11}+n; c_{10}+1; c_3 s) \tag{A16}$$

where $c_{12} > 0$, $c_{13} > 0$ and $s \in [0, 1/c_3]$, $c_3 \neq 0$.

**Appendix B: Calculation of the Normalization Constant in the SS equation**

To compute the normalization constant $A_{nl}$ in Eq. (19), it is easy to show with the use of $R_{nl}(r) = r^{-1} \psi_{nl}(r)$, that

$$\int_0^\infty |R_{nl}(r)|^2 r^2 dr = \int_0^\infty |\psi_{nl}(r)|^2 dr = \int_0^1 |\psi_{nl}(s)|^2 \frac{ds}{2\alpha s} = 1, \tag{B1}$$

where we have used the substitution $s = e^{-2ar}$. Inserting Eq. (18) into (B1) and using the following definition of the Jacobi polynomials [59]

$$P_n^{(p_0, w_0)}(1 - 2s) = \frac{\Gamma(n + p_0 + 1)}{n! \Gamma(p_0 + 1)} {}_2F_1(-n, p_0 + w_0 + n + 1; 1 + p_0; s), \tag{B2}$$

where $p_0 = 2\varepsilon_{nl}$ and $w_0 = 2\nu$, we arrive at

$$|A_{nl}|^2 \int_0^1 s^{2\varepsilon_{nl}-1}(1-s)^{2\nu+1}\left\{{}_2F_1(-n, 2\varepsilon_{nl}+2\nu+1+n; 1+2\varepsilon_{nl}; s)\right\}^2 ds$$

$$= 2a\left(\frac{n!\Gamma(2\varepsilon_{nl}+1)}{\Gamma(n+2\varepsilon_{nl}+1)}\right)^2, \tag{B3}$$

where ${}_2F_1$ is the hypergeometric function. Using, the following integral formula [60,61]

$$\int_0^1 z^{2\lambda-1}(1-z)^{2(\eta+1)}\left\{{}_2F_1(-n, 2(\lambda+\eta+1)+n; 1+2\lambda+1; z)\right\}^2 dz$$

$$= \frac{(n+\eta+1)n!\Gamma(n+2\eta+2)\Gamma(2\lambda)\Gamma(2\lambda+1)}{(n+\eta+\lambda+1)\Gamma(n+2\lambda+1)\Gamma(n+2\lambda++2\eta+2)}, \quad \eta > -\frac{3}{2}, \quad \lambda > 0, \tag{B4}$$

we can get the normalization constant as

$$A_{nl} = \sqrt{\frac{2a\varepsilon_{nl}n!(2n+2\nu+2\varepsilon_{nl}+1)\Gamma(n+2\nu+2\varepsilon_{nl}+1)}{(n+\nu+1/2)\Gamma(n+2\nu+1)\Gamma(n+2\varepsilon_{nl}+1)}}. \tag{B5}$$

The relation (B5) can be used to compute the normalization constant for $n = 0, 1, 2, \cdots$. for the ground state ($n = 0$), we have

$$A_{0l} = \sqrt{\frac{a(2\nu+2\varepsilon_{nl}+1)}{(\nu+1/2)B(2\varepsilon_{nl}, 2\nu+1)}}, \tag{B6}$$

where

$$B(2\varepsilon_{nl}, 2\nu+1) = \frac{\Gamma(2\nu+1)\Gamma(1+2\varepsilon_{nl})}{2\varepsilon_{nl}\Gamma(2\nu+1+2\varepsilon_{nl})}. \tag{B7}$$

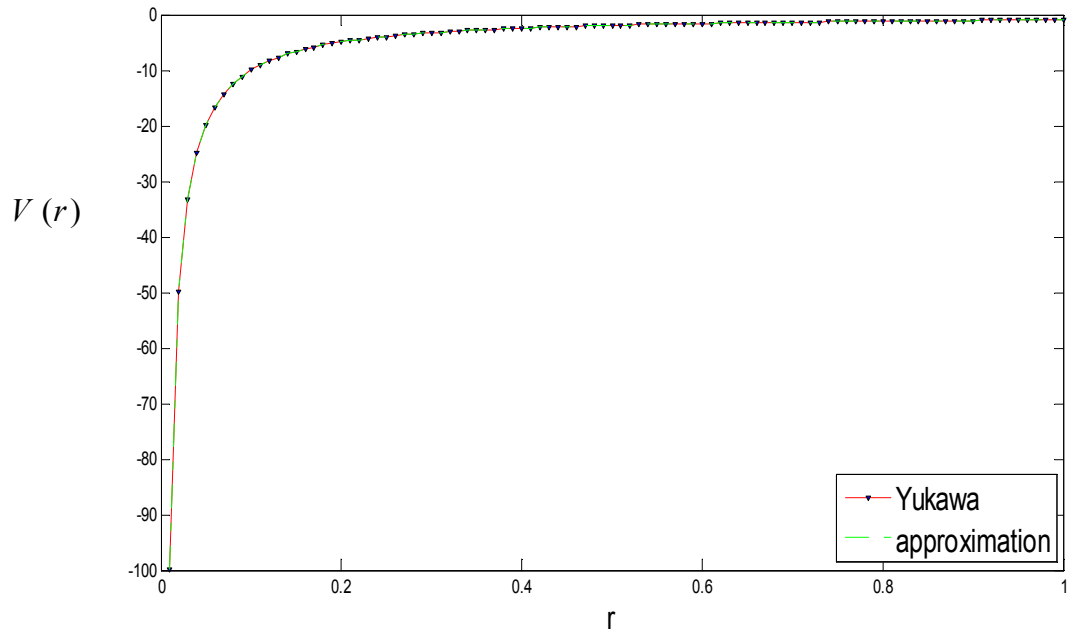

**Figure 1:** Yukawa potential (1) (red curve) and its approximation (8) (green curve).

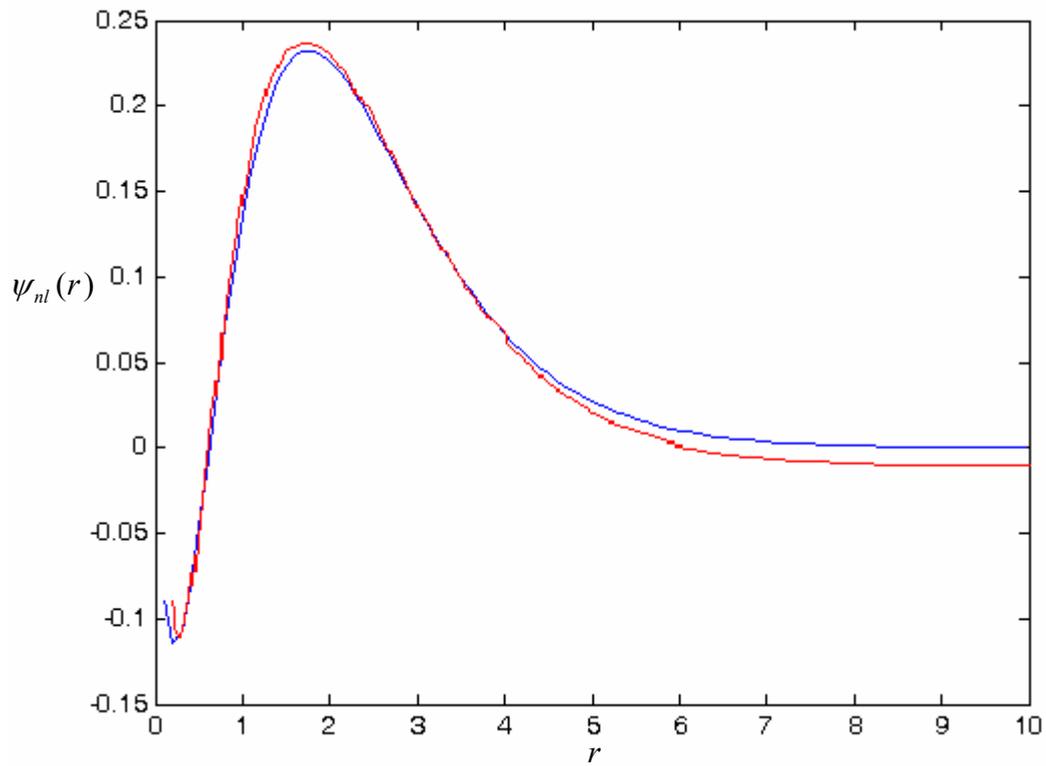

**Figure 2.** Approximation (red curve) and exact numerical (blue curve) wave functions of the SS equation with the Yukawa potential for $n=1$ and $l=0$.

**Table 1** Approximation and numerical binding energy spectrum of the SS equation with the Yukawa potential for various values of $n$ and $l$ quantum numbers.

| $n$ | $l$ | $-E_{nl}$ | | | | | |
|---|---|---|---|---|---|---|---|
| | | $a = 0.01 fm^{-1}$ | | $a = 0.005 fm^{-1}$ | | $a = 0.001 fm^{-1}$ | |
| | | Approximation | Numerical | Approximation | Numerical | Approximation | Numerical |
| 1 | 0 | 0.5032 | 0.5035 | 0.5082 | 0.5084 | 0.5122 | 0.5124 |
| 2 | 0 | 0.1843 | 0.1912 | 0.1892 | 0.1960 | 0.1932 | 0.2000 |
|   | 1 | 0.0709 | 0.0718 | 0.07568 | 0.0765 | 0.0796 | 0.0803 |
| 3 | 0 | 0.0907 | 0.0978 | 0.0956 | 0.1025 | 0.0995 | 0.1064 |
|   | 1 | 0.0419 | 0.0416 | 0.0465 | 0.0460 | 0.0504 | 0.0499 |
|   | 2 | 0.0258 | 0.0258 | 0.0303 | 0.0300 | 0.0341 | 0.0337 |
| 4 | 0 | 0.0516 | 0.0568 | 0.0563 | 0.0615 | 0.0602 | 0.0653 |
|   | 1 | 0.0262 | 0.0266 | 0.0307 | 0.0308 | 0.0345 | 0.0346 |
|   | 2 | 0.0167 | 0.0165 | 0.0210 | 0.0204 | 0.0248 | 0.0241 |
|   | 3 | 0.0109 | 0.0116 | 0.0149 | 0.0153 | 0.0187 | 0.0189 |
| 5 | 0 | 0.0317 | 0.0355 | 0.0362 | 0.0399 | 0.0401 | 0.0437 |
| 6 | 0 | 0.0203 | 0.0231 | 0.0247 | 0.0273 | 0.0285 | 0.0311 |
| 7 | 0 | 0.0133 | 0.0154 | 0.0174 | 0.0194 | 0.0211 | 0.0231 |

**Table 2** Approximation and numerical binding energy spectrum of the Schrödinger equation with the Yukawa potential for various values of $n$ and $l$ quantum numbers.

| $n$ | $l$ | $-E_{nl}$ | | | | | |
|---|---|---|---|---|---|---|---|
| | | $a = 0.01 fm^{-1}$ | | $a = 0.005 fm^{-1}$ | | $a = 0.001 fm^{-1}$ | |
| | | Approximation | Numerical | Approximation | Numerical | Approximation | Numerical |
| 1 | 0 | 0.2926 | 0.2944 | 0.3025 | 0.3019 | 0.3105 | 0.3103 |
| 2 | 0 | 0.1191 | 0.1229 | 0.1289 | 0.1287 | 0.1369 | 0.1342 |
|   | 1 | 0.0584 | 0.0577 | 0.0682 | 0.0684 | 0.0761 | 0.0762 |
| 3 | 0 | 0.0584 | 0.0680 | 0.0682 | 0.0731 | 0.0761 | 0.0778 |
|   | 1 | 0.0305 | 0.0334 | 0.0401 | 0.0413 | 0.0480 | 0.0475 |
|   | 2 | 0.0154 | 0.0155 | 0.0249 | 0.0242 | 0.0327 | 0.0320 |
| 4 | 0 | 0.0305 | 0.0419 | 0.0401 | 0.0467 | 0.0480 | 0.0509 |
|   | 1 | 0.0154 | 0.0214 | 0.0249 | 0.0277 | 0.0327 | 0.0330 |
|   | 2 | 0.0065 | 0.0095 | 0.0157 | 0.0165 | 0.0235 | 0.0229 |
|   | 3 | 0.0008 | 0.0010 | 0.0098 | 0.0099 | 0.0175 | 0.0178 |
| 5 | 0 | 0.0154 | 0.0274 | 0.0249 | 0.0318 | 0.0327 | 0.0359 |
| 6 | 0 | 0.0065 | 0.0084 | 0.0157 | 0.0226 | 0.0235 | 0.0264 |
| 7 | 0 | 0.0008 | 0.00105 | 0.00985 | 0.0165 | 0.0175 | 0.0202 |